# Viral Hitchhikers and Macroevolution: A Novel Hypothesis on Explosive Speciation

Mario E. Lozano[1,2]; Marcela G. Pilloff[1,3]

1: Laboratorio de Virus emergentes, Instituto de Microbiología Básica y Aplicada, Universidad Nacional de Quilmes

2: Consejo Nacional de Investigaciones Científicas y Técnicas (CONICET)

3. Instituto de Biotecnología, Universidad Nacional de Hurlingham

ORCID Numbers:

Mario Lozano: 0000-0002-5855-0932

Marcela Pilloff: 0000-0002-0429-9869

*Correspondence: Mario Lozano

e-mail: mario.lozano@unq.edu.ar

X: @LozanoMarioE

BSKY: @mariolozano.bsky.social

## Data Availability

No new data were generated for this theoretical study.

## Keywords

"K-Pg mass extinction", "punctuated equilibrium", "EVEs", "adaptive radiation" "Speciation"

**Abstract**

Mobile genetic elements (e.g., endogenous viruses, LINEs, SINEs) can transfer between genomes, even between species, triggering dramatic genetic changes. Endogenous viral elements (EVEs) arise when infectious viruses integrate into the host germline. EVEs integrate at specific sites; their genes or regulatory regions can be exapted and could induce chromosomal rearrangement. We propose that EVEs participate in adaptive radiations and that their parent viruses, through interspecific transfer, could initiate new species formation. By synchronously affecting multiple individuals, viral outbreaks generate shared genomic changes that both facilitate reproductive isolation and provide the simultaneous modifications needed for groups to emerge as founders of new species. We suggest horizontal viral transfer during the K-Pg accelerated mammalian radiation linking viral epidemics to macroevolutionary diversification.

**The Puzzle of Explosive Speciation**

Adaptive radiations challenge gradualist models of speciation. While ecological factors (e.g., vacant niches) and sexual selection are well-studied, the genomic triggers of rapid diversification remain elusive. Horizontal transfer of **endogenous viral elements** (EVEs, see Glossary) could explain abrupt speciation patterns [1], complementing classical adaptive radiation models [2]. EVEs transferred genes may have a viral or host origin.

Mounting evidence implicates mobile genetic elements as drivers of genomic changes underlying phenotypic divergence. Mobile genetic elements are molecular entities that can transpose inside genomes of cellular organisms or between cells. Transposable elements (TEs), once dismissed as "trash DNA," are ubiquitous in eukaryotic genomes and frequently exapted to generate novel coding sequences, non-coding RNAs, and regulatory elements. As detailed in [3], TEs integrate at specific genomic loci where they: (i) provide new genes or RNAs (ii) induce mutations and chromosomal rearrangements, (iii) modulate gene expression and immune responses, and (iv) affect both somatic and germline cells. Their integration has facilitated complex vertebrate traits, including the nervous system, adaptive immunity, and placenta [4]. Retroelements are a major class of TEs that move within the genome via an RNA intermediate, using a "copy-and-paste" mechanism, by means of reverse transcription. Most retroelements (LINEs, SINEs, ERVs) share evolutionary ties with ancient viruses, either as direct viral descendants or through *trans*-acting viral dependencies. Notably, SINEs

> **Highlights**
>
> 1. Endogenous viral elements (EVEs), remnants of viruses integrated into host genomes, are potent drivers of genomic novelty, promoting chromosomal rearrangements, gene exaptations, and reproductive isolation.
>
> 2. EVEs are not merely "molecular fossils": their capacity to modulate regulatory networks and generate evolutionary novelties (e.g., syncytins in placenta) reveals their key role in adaptive radiations.
>
> 3. Horizontal viral transfer between species, particularly during ecological crises, could explain patterns of abrupt speciation by inducing massive and synchronous genomic changes across populations. Ancestral viral outbreaks may have acted as triggers for punctuated speciation.
>
> 4. The hypothesis integrates virology, paleogenomics, and evolutionary theory, proposing viruses as macroevolutionary agents during environmental crisis, with implications for understanding rapid post-K-Pg diversification.

have been linked to rapid cichlid fish radiations via *egg-spot* regulation modifications [2]—a clade diversifying into ~2,000 species within <10 Myr.

The post-K-Pg boreoeutherian mammal radiation (~65 Ma) represents a pivotal evolutionary event, with more than 90% of mammalian orders emerging within ~10 Myr. Most competing models (explosive, soft explosive, trans-KPg, long/short fuse [5]) agree that peak diversification coincided with the extinction event. While ecological vacancy after dinosaur extinction is often cited as the primary driver, the molecular mechanisms enabling this explosive speciation remain unclear. The rapid diversification of post-K-Pg mammals defies gradualist evolutionary models. We propose that horizontal transfer of EVEs and other mobile elements from dinosaur carcasses to ancestral mammals facilitated: (i) large-scale genomic rearrangements, (ii) metabolic adaptations, (iii) gene exaptations, and (iv) reproductive isolation—key speciation factors.

For speciation to occur, these processes must meet two conditions:

1. Introduce substantial genomic changes, as mentioned
2. Synchronously affect multiple individuals (sharing both time and space) overcoming fitness barriers.

**Core hypothesis**: EVEs acquired via horizontal transfer during mass extinctions (e.g., K-Pg) triggered rapid speciation through:

- Abrupt genomic changes and genome rearrangements inducing reproductive isolation
- Modified host-virus interactions that amplified selectable variation via cross-species jumps (e.g., dinosaurs→mammals)

This hypothesis integrates evolution, virology and paleogenomics, offering testable predictions about how viral transfer during ecological crises accelerated mammalian speciation.

**EVEs as Drivers of Genomic Innovation**

Multiple viral families have colonized vertebrate genomes since the Mesozoic [6-8] (Table 1). Endogenization events often coincide with adaptive radiations (e.g., Retroviridae in post-K-Pg mammals [7]). Endogenous retroviruses (ERVs) are the largest group of EVEs seen in mammalian genomics [9]. Retroviruses, RNA viruses confined to vertebrates, have genomes with long terminal repeat (LTR) sequences as well as three core genes (gag, pol, env). They replicate by reverse transcription and integrating into the host genome. When retroviruses infect germline cells, the integrated proviruses can be vertically inherited, forming ERVs—which likely gave rise to other mobile elements like LINEs and LTR retrotransposons.

Table 1. Endogenized viral families in vertebrates and their evolutionary impact

| Viral Family | Host range | Insertion time (Ma) | References |
|---|---|---|---|
| **Retroviridae** | Mammals, birds | >39 | [12,37] |
| **Filoviridae** | Bats, primates, marsupials, cartilaginous fish, Komodo dragon | >30 | [8,41] |
| **Bornaviridae** | Primates, afrotherians, lemurs | >93 | [42,19] |
| **Parvoviridae** | Mammals | >30 | [43,44] |
| **Circoviridae** | Carnivores (*cat, dog, panda*), birds, reptiles, amphibians, fish | >68 | [19,45] |
| **Hepadnaviridae** | Passerine birds | 19-80 | [7,46-47] |
| **Hepacivirus** | Rodents | >14 | [48] |
| **Nairoviridae** | Etruscan shrew | >5 | [48] |
| **Benyviridae** | Amphibians, sharks, reptiles | >50 | [48] |
| **Paramyxoviridae** | Teleost fish | >11 | [48] |
| **Chuwiridae** | Teleost fish, marsupials | >12 | [48] |
| **Herpesviridae** | Primates | >60 | [49] |

ND: No data. Host ranges based on endogenous viral elements (EVEs) confirmed via paleovirology.

Most ERVs become molecular fossils through disabling mutations, preserving records of virus-host coevolution. In humans, ERV-derived sequences comprise more than 8% of the genome, with LINEs (17%), non-ERV LTR retrotransposons (2%), and non-retroviral EVEs (1%) demonstrating their pervasive influence [10]. Unlike other viruses, integrated ERVs propagate via two distinct mechanisms:

1. **Reinfection**: Functional ERVs produce viral particles to infect new germ cells. Defective ERVs (or other EVEs) without a complete set of core genes, can achieve this through *trans*-complementation by co-infecting functional retroviruses [9].

2. **Retrotransposition**: ERVs amplify via RNA intermediates that are reverse-transcribed and re-integrated, either using their own proteins (*cis*-mobilization) or hijacking other TE's machinery (*trans*-complementation) [11]. Most retroelements replicate by retrotransposition, via trans-complementation using ERV´s derived proteins.

The dependence of other viruses on retroelement-encoded proteins may explain ERVs' genomic dominance. New ERV insertions undergo genetic drift and are subject to natural selection, resulting in insertion polymorphisms that are later fixed or disappears. Notably, 412 recently active ERVs (ERVi) were identified across 123 vertebrates [12], revealing ongoing retroviral colonization—a continuum where ancient ERVi become modern ERVs.

**EVEs as Speciation Catalysts.**

The coevolution of ERVs and other EVEs with vertebrate genomes is becoming increasingly evident. But what if the incorporation of these viral genomes, and the proliferation of their descendant elements (ERVs or EVEs) within host genomes, induced profound genomic changes that aided speciation?. Three properties position EVEs as potent drivers of speciation in animals:

1. **Genetic novelty generation**: EVEs preferentially integrate near developmentally critical genes (Lu, Jia, PhD thesis, University of California San Diego, 2019), where they can:

    - Provide regulatory elements (e.g., a primate-specific ERVP71A-LTR as an enhancer for placental HLA-G [13])
    - Modify gene networks via epigenetic silencing [14] or new enhancers [15]
    - Contribute exapted exons and promoters as the Fv1 antiretroviral restriction factor in mouse derived from MERV-L [16].

2. **Reproductive isolation**: Site-specific integrations can induce chromosomal rearrangements or inversions in the parent's germline cells that reproductively isolate their offspring populations [17].

3. **Synchronized population modification**: During viral outbreaks most infected individuals can undergo identical genetic modifications at similar times. Crucially, when infectious virus:
    - Target conserved integration sites, generating parallel mutations
    - Affect both sexes with compatible genomic changes
    - Confer adaptive advantages for new generations in new niches

This combination satisfies the rare conditions needed for "instantaneous" speciation—bypassing the fitness valley typically associated with major genomic changes in single individuals. After the emergence of a potential new species, at least two individuals, one male and one female, with the same genetic modifications in the same geographic area and time period, are required to ensure their reproduction. K-Pg mass extinction may have created ideal conditions for such viral-mediated speciation events in early mammals.

**A Cross-Class Trigger: Sauropsid-Mammal Viral Transfer and Macroevolution.**

Fossilized viral elements reveal an ancient association between mammals and viruses dating to their cladogenesis. Orthologous EVEs shared across species provide minimum estimates of host divergence times [18]. Notable examples include:

- Bornavirus-derived elements in boreoeutherians (>100 Ma) [19]
- Foamy viruses in vertebrates (>400 Ma) [20]
- Lagomorph-specific ERV, RELIK (>12 Ma) [6]
- Lemur ERV, PSIV (>10 Ma) [6]

Strikingly, younger EVE integration events coincide with post-K-Pg mammalian radiations, suggesting their potential speciation role. We propose that dinosaur-origin viruses infected boreoeutherian ancestors, enabling cross-species genetic exchange and inducing significative phenotypic variation.

**The K-Pg Transmission Window.** In the context of the K-Pg mass extinction, numerous dying or dead dinosaurs may have been scavenged by small, omnivorous animals in search of a scarce food (Figure 1A). Normally limited by ecological separation, dinosaur viruses (retroviruses and others) present in blood, viscera, and other internal tissues, became accessible to scavenging mammals for the first time.

When these dinosaur viruses successfully infected mammalian hosts, they gained the potential to integrate into germline genomes. The initial integration event represented just the first step in a cascade of unlikely but biologically plausible processes: the primary infected individual could then transmit the virus to nearby conspecifics, potentially creating localized outbreaks. For these viral elements to drive speciation, several subsequent conditions needed to be met simultaneously - the viruses had to integrate into germ cells of multiple individuals from both genders, target specific genomic sites capable of inducing chromosomal rearrangements, and provide exaptable genetic material that could generate new metabolic capabilities or regulatory elements.

Critically, this process would have varied across different habitats during the K-Pg extinction. Different combinations of dinosaur, viruses and local mammalian populations could have produced varied genomic outcomes, potentially generating even greater initial diversity than currently observed. Finally, some populations emerging under these conditions adapted better to their environments and prevailed, giving rise to the current diversity of boreoeutherians. This scenario finds parallels in documented cases of horizontal viral transfer among contemporary species, such as the gammaretrovirus endogenization events shared by bat, feline and pangolin lineages 13-25 million years ago [21].

**Mechanistic Evidence: From ERVs to Speciation.** Additional evidence comes from the RnERV-K8e ERV family in rats, whose recent activity has generated insertional polymorphisms among inbred strains, as observed in [22]. These elements not only alter gene expression through intronic or exonic insertions (e.g., the *Cntrob* gene) but can also transfer between closely related species like rats and mice, facilitating genetic divergence. Interestingly, centrobina—the protein encoded by *Cntrob*—is a centriole-associated protein asymmetrically localized to daughter centrioles and required for centriole duplication and cytokinesis. Its modification, mediated by RnERV-K8e insertion, could induce chromosomal rearrangements (Figure 1B). The presence of autonomous functional copies of this retroelement, such as Rat-ρ on rat chromosome 20, suggests ERVs act as engines of genomic innovation during speciation, whether through disruptive mutations, splice site modifications, or regulatory sequence introductions. Finally, these authors show that a member of the RnERV-K8e retrovirus family may have reactivated in rats, transmitted to mice, and evolved into a new endogenous retroelement family (NT_039515.6) in this new host. Thus, ERVs emerge as dynamic evolutionary agents capable of transferring between species, shaping genomes, and potentially influencing reproductive isolation between populations.

**Syncytins: A Paradigm of Viral Exaptation.** Beyond insertional polymorphisms, another striking manifestation of viral-mediated evolution is seen in syncytins, fusogenic proteins derived from retroviral envelope (env) glycoproteins of ERVs, essential for placental formation in mammals [23]. Placental function depends on cell fusion, a phenotypic trait that in some animals is mediated by syncytin. The conservation of open reading frames in syncytin genes across mammalian species, coupled with low mutation rates and minimal polymorphism, underscores their critical role in mammalian evolution. Most elements involved in syncytin production descend from betaretroviruses, with the exception of endogenous gammaretroviruses in mouse, mole rat, jerboa, vole, and hamster genomes, integrated ~40 million years ago. Betaretroviruses in other rodents (rats, guinea pigs) integrated ~20 million years ago. Two endogenous retroviruses are responsible for syncytin expression in primates, HERV-W (found in humans, chimpanzees, gorillas, and other Old World monkeys) and HERV-FRD (found in the same species mentioned above and orangutans). HERV-W has integrated ~25 million years ago during catarrhine radiation, and HERV-FRD ~40 million years ago during hominoid-cercopithecid divergence.

Lagomorphs (rabbits, hares) acquired OryERV ~12 million years ago [24]; carnivores (dog, felines) integrated Syncytin producing ERV ~40 million years ago [25-26]; whereas higher ruminants (artiodactyls) have the BovER-K gene integrated over 40 million years ago [23]. Syncytins are absent in other mammalian orders (Chiroptera, Eulipotyphla, Cetacea, Perissodactyla, Xenarthra, Pholidota, Tubulidentata, Macroscelidea, Dermoptera, Scandentia), which rely on other fusogens for placental function. Indeed, both basal (e.g., Xenarthra) and highly specialized (e.g., Cetacea) mammalian orders lack syncytins. These findings suggest that syncytin domestication took place independently in different mammalian lineages after multiple individual retroviral element capture events to facilitate placental formation (Figure 1C). This emphasizes the nature of retroviruses as an inherent contributor to mammalian diversity.

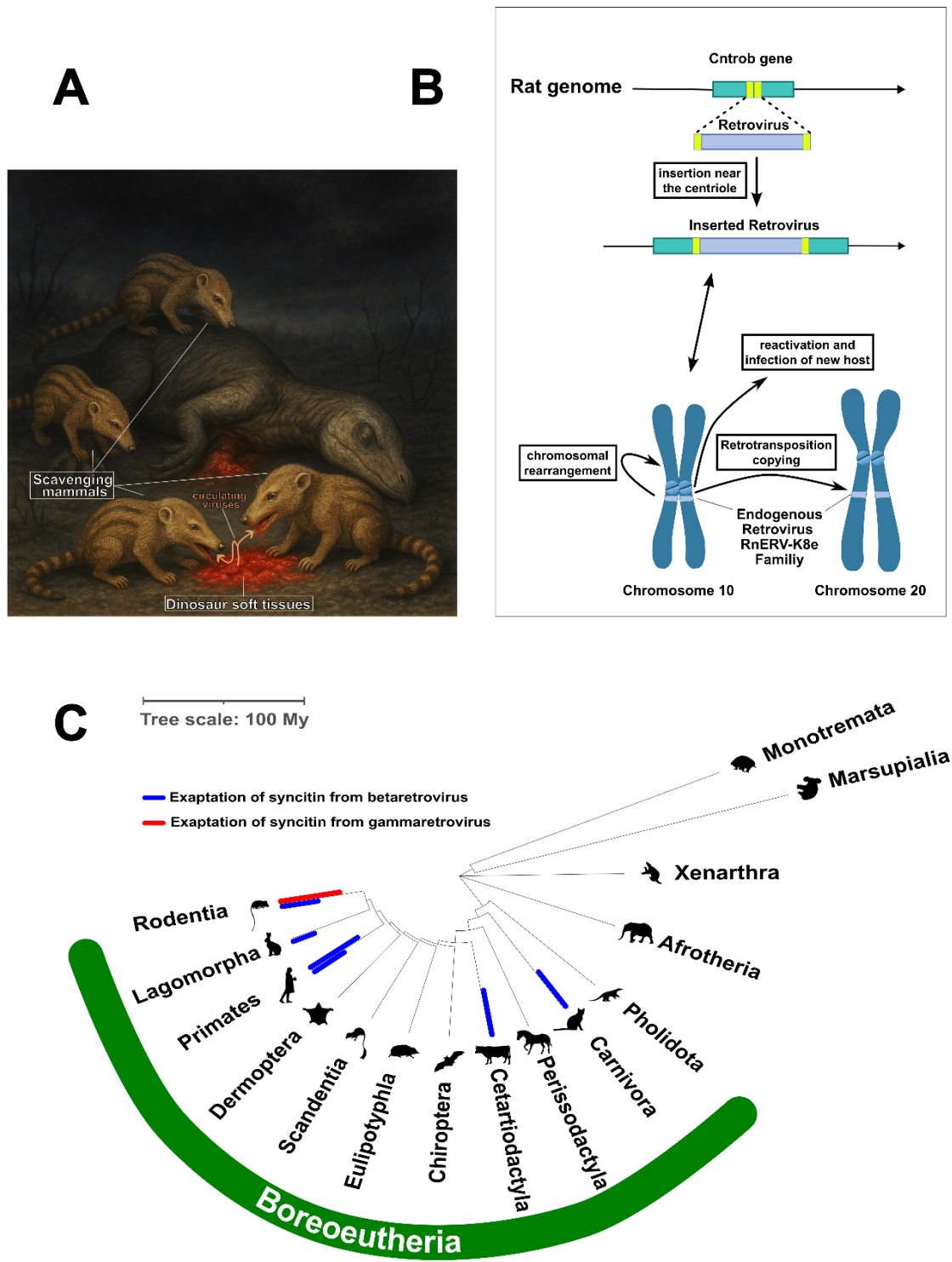

Figure 1: Proposed mechanisms of viral transfer and speciation. A. Post-K-Pg scenario. Artist's impression generated by artificial intelligence using DALL·E at *https://openai.com*, based on specific descriptions by the author. It illustrates a possible interaction between small Mesozoic mammals and a dying dinosaur during the K-Pg extinction event. The scene shows the scavenging mammals feeding on soft tissue, raising a plausible scenario for the zoonotic transfer of viruses or other pathogens from dinosaurs to surviving mammalian lineages, against a backdrop of ecological collapse, species overcrowding, and widespread physiological stress. The image is for illustrative purposes and does not represent direct evidence, but rather a hypothesis based on ecological and virological principles. B. RnERV-K8e dynamics in Rattus. The insertion of the retrovirus (RnERV-K8e) near the centromeric region of chromosome 10 of R. norvegicus, possible chromosomal rearrangements and mechanisms of viral dispersal by (i) retrotransposition to new

loci (e.g. chromosome 20) and (ii) reactivation of the virus to infect new hosts are shown. Based on data from [22]. C. Phylogeny of mammals with syncytin integration events. Branch lengths are related to divergence times in millions of years (Ma). Thick lines parallel to some branches of the tree indicate the estimated time of entry into the genome of that clade of the retroviruses that contributed the syncytins that induce placenta formation (in blue, betaretroviruses and in red, gammaretroviruses). Time-scaled phylogeny (top left panel and bottom bar). Phylogenetic tree generated and visualized with iTOL v5 [50].

**Not only retrovirus.** Bornavirus-like elements (EBLs) provide additional evidence, with primate EBL integrations coinciding with K-Pg. Endogenous bornavirus-like elements (EBLs) are the most abundant RNA virus-derived fossils in vertebrate genomes [7,27] and have been detected in a wide variety of animals, including all placental mammal clades (except Xenarthra), marsupials, birds, and reptiles [19]. During their replication cycle, bornaviruses produce only RNA as replication intermediates, necessitating trans-acting elements (e.g., reverse transcriptase or integrase) derived from retroelements (retrotransposons or ERVs) for genome integration. A carbovirus-like EBL, named EBLN23, detected in bird and reptile genomes and estimated to be over 66 million years old, is evolutionarily related to other carbovirus-like elements—EBLN10, EBLN11, EBLN15, EBLN16, and EBLN17—which intriguingly endogenized in primate genomes around the K-Pg extinction event [19,28].

**Concluding remarks: EVEs as Macroevolutionary Architects.**

The proposed hypothesis suggests that endogenous viruses (EVEs), acquired via horizontal transfer from dinosaurs to boreoeutherian mammals, acted as genomic catalysts for explosive post-K-Pg speciation (Figure 2). This model integrates three key pillars:

1. **Genomic disruptive capacity of EVEs**: These elements could drive rapid chromosomal rearrangements, regulatory novelties, and reproductive isolation over short timescales

2. **Epidemiological synchrony**: Viral outbreaks enable population-wide genomic changes, overcoming fitness bottlenecks limiting fixation of isolated mutations

3. **Unique ecological context**: Dinosaur carrion during the K-Pg event facilitated cross-species viral jumps—a rare phenomenon under normal conditions.

Though speculative, the hypothesis offers testable predictions:

- Comparative genomics of EVEs in birds and boreoeutherian mammals could test this hypothesis, with implications for understanding ancient epidemics and speciation.
- Paleovirological analyses of Paleocene mammalian fossil genomes might identify viral insertions associated with cladogenesis.
- Bioinformatics simulations of EVE integration into key developmental genes.

These predictions, along with key unresolved questions about the role of viruses in macroevolution, are highlighted in the Outstanding Questions box.

We conceptualize speciation not as a gradual or analog process but rather as an instantaneous, abrupt genomic modification akin to a quantum leap. If validated, this hypothesis could redefine our understanding of adaptive radiations, positioning viruses as agents of "punctuated speciation" (sensu Gould & Eldredge [29]), where mass infectious events trigger evolutionary jumps.

Our model extends Margulis' paradigm of symbiogenesis [30], suggesting transient viral interactions—without stable symbiosis—can also induce speciation. While Margulis emphasized cellular cooperation, EVEs would act as "genomic agitators" during ecological crises as described [31,32]. Margulis noted that the deepest speciation event in evolutionary history was likely eukaryogenesis—a unique, abrupt event driven by massive gene co-option from symbionts [33]. The first eukaryote was a genomic hybrid with archaeal and bacterial contributions, explaining the chimeric nature of modern eukaryotic genomes [34-35]. This large-scale co-option and exaptation of genetic material and biochemical processes resulted not merely in a new species, but in an entirely novel domain of life.

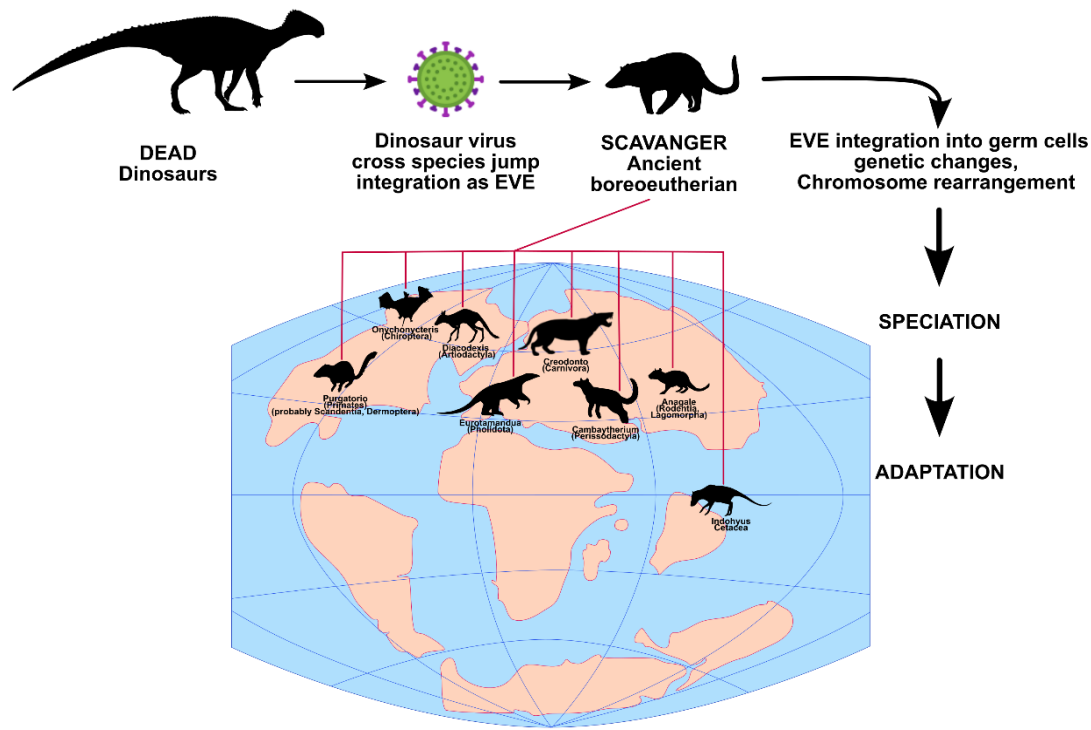

Figure 2. Endogenous viruses as drivers of speciation in post-K-Pg mammals. The hypothesis of viruses as speciation triggers in post-K-Pg mammals is illustrated, including the origin of genetic variation (dinosaurs), vectors (viruses), recipients (ancestral mammals), and modern mammalian orders located in the ancient Earth geography, where they are presumed to have arisen [51]. The processes required for speciation are listed at the top of the figure. All silhouettes were obtained from PhyloPic, free silhouette images of organisms (version 2.0).

At a much smaller scale, while viruses are unlikely to generate new domains or kingdoms (given their limited gene transfer capacity compared to bacteria/archaea), they are fully equipped to accompany new species creation. Speciation requires a new population directly descended from another (its direct parents), separated by significant genetic differences—whether chromosomal number/organization (as in King's chromosomal speciation models [36]) or metabolic system presence/scale (e.g., loss of ascorbic acid synthesis in primates [37]). These differences render populations genetically incompatible and phenotypically distinct. These changes must simultaneously affect both sexes and generate identical genetic modifications in all members of the nascent generation to transform them into a new species. Unlike classical Dobzhansky-Muller incompatibility models—which require gradual mutation accumulation in isolated populations—viruses could induce massive, synchronous genomic changes in a single generation, as suggested by recent ERV-K activity in humans [38] and SINE-associated rapid speciation in African cichlids [2]. For the first time in history, a group of parents could have offspring constituting not just a new generation but an entirely new species. The change is not gradual but quantum. Viruses, as we propose, represent the most plausible vectors for this abrupt, synchronous change.

Although the synchrony of virus-induced genomic changes might appear improbable, viral outbreaks in small, isolated populations (like post-K/Pg mammals) could facilitate it, as suggested by insertional polymorphisms observed in RnERV-K8e [22]. Notably, syncytin-like exaptations have been identified beyond mammals, including viviparous reptiles [39], suggesting viral domestication may be a universal mechanism for reproductive innovations during rapid radiations.

This model links paleovirology [6,18] with symbiogenesis [30] and punctuated equilibrium theory [29], positioning ancestral viral outbreaks as catalysts for adaptive radiations. It expands viruses' ecological role beyond pathogenicity, revealing them as underestimated macroevolutionary drivers and biodiversity creators during planetary crises. Aligned with mammalian radiation chronology, the hypothesis offers testable predictions through comparative genomics of extant species. Future studies analyzing selection signatures in EVEs [40] shared across distant lineages (e.g., birds/reptiles/mammals) could test this hypothesis.

## Outstanding Questions:

**1. How conserved are EVE integration patterns across avian, reptilian and mammalian lineages?**
Comparative genomic analyses of EVEs shared between sauropsids (birds/reptiles) and boreoeutherians could reveal horizontal transfer events during the K-Pg extinction, testing the dinosaur-mammal viral transmission hypothesis.

**2. Can paleovirology detect extinct viral elements in Mesozoic mammalian fossils?**
Advances in ancient DNA recovery from Paleocene fossils may identify ERV integration events linked to post-K-Pg cladogenesis, offering direct evidence of viral-driven speciation.

**3. Do EVEs preferentially target developmental genes during cross-species transmission?**
Experimental integration of reconstructed Mesozoic-like retroviruses into model organisms (e.g., mice) could test whether EVEs disrupt conserved regulatory networks, driving reproductive isolation.

**4. How do viral outbreaks synchronize speciation-compatible genomic changes in populations?**
Epidemiological models simulating retroviral spread in post-extinction mammal populations could quantify the likelihood of parallel germline integrations overcoming fitness barriers.

**5. Are syncytin-like exaptations a recurring mechanism in rapid radiations?**
The convergent domestication of retroviral envelope proteins in both mammals (*syncytins*) and viviparous reptiles (e.g., *Mabuya* lizards) suggests this mechanism could be a widespread driver of reproductive innovation. Screening non-mammalian adaptive radiations (e.g., cichlids, fish, amphibians) may reveal if viral gene co-option is a fundamental speciation trigger.

**Acknowledgement**

This work was supported by the Universidad Nacional de Quilmes, Argentina. MEL is titular professor of UNQ and member of CONICET research career, MGP is associated professor of UNAHUR. The authors would like to thank Victor Romanowski, Mariano Belaich, Diego Golombek and Antonio Lagares (s) for their invaluable comments and suggestions.

**Ethics Statement**

Not applicable: This work is theoretical and did not involve human or animal experimentation.

**Declaration of interests**

The authors declare no competing interests.

## Glossary:

**Adaptive radiation**: Rapid diversification of a single ancestral lineage into multiple species, often driven by ecological opportunity (e.g., vacant niches).

**Endogenous viral elements (EVEs)**: Viral-derived DNA sequences integrated into host genomes, including retroviruses (ERVs) and non-retroviral elements.

**Exaptation**: Evolutionary process where a trait (e.g., a viral protein) is co-opted for a new function unrelated to its original role. Example: Syncytins, retroviral envelope proteins repurposed for placental development.

**Germline cells**: Reproductive cells (sperm, eggs) that transmit genetic information to offspring.

**Gradualist models**: Evolutionary theories emphasizing slow, incremental changes as the primary driver of speciation, contrasting with punctuated equilibrium.

**Horizontal gene transfer**: Acquisition of genetic material from unrelated organisms, contrasting with vertical inheritance.

**K-Pg mass extinction**: Cataclysmic event ~66 million years ago, marked by an asteroid impact and volcanic activity, causing the extinction of non-avian dinosaurs and ~75% of Earth's species, followed by mammalian adaptive radiation.

**Long terminal repeats (LTRs)**: DNA sequences flanking retroviral genomes, essential for integration and replication.

**Macroevolution**: Large-scale evolutionary changes over geologic time, including speciation patterns, extinction events, and the emergence of novel traits.

**Outbreaks**: Sudden, widespread occurrence of infectious disease within a population. In evolutionary contexts, viral outbreaks could synchronize genomic changes across multiple individuals, facilitating speciation.

**Paleogenomics**: Study of ancient DNA from fossils or historical specimens to reconstruct evolutionary processes and genomic changes.

**Punctuated equilibrium**: Evolutionary model proposing long periods of stasis interrupted by rapid speciation events.

**Reproductive isolation**: Biological barriers preventing interbreeding between populations, a hallmark of speciation.

**Reverse transcription:** Enzymatic process, part of the expanded central dogma of molecular biology, by which DNA is synthesized from an RNA template, catalyzed by the enzyme reverse transcriptase.

**Symbiogenesis**: Evolutionary theory emphasizing symbiosis as a driver of major innovations (e.g., mitochondria).

**Syncytins**: Retroviral envelope proteins exapted for placental cell fusion in mammals and some reptiles (e.g., *Mabuya* lizards).

**Transposable elements (TEs)**: Mobile DNA sequences capable of changing position within genomes, often contributing to genetic novelty.